\documentclass[twocolumn]{aastex62}

\usepackage{enumitem}
\usepackage{amsmath}

\shorttitle{Magnetic Activity of 53 M-Dwarfs}
\shortauthors{Boudreaux et al.}

\begin{document}

\title{The Ca II H\&K Rotation-Activity Relation in 53 mid-to-late type M-Dwarfs}

\author[0000-0002-2600-7513]{Emily M. Boudreaux}
\affiliation{Department of Physics and Astronomy, Dartmouth College, Hanover, NH 03755, USA}

\author[0000-0003-4150-841X]{Elisabeth R. Newton}
\affiliation{Department of Physics and Astronomy, Dartmouth College, Hanover, NH 03755, USA}

\author[0000-0003-0295-293X]{Nicholas Mondrik}
\affiliation{Department of Physics , Harvard University , 17 Oxford Street Cambridge MA 02138, USA}

\author[0000-0002-9003-484X]{David Charbonneau}
\affiliation{Center for Astrophysics, Harvard \& Smithsonian, 60 Garden Street, Cambridge, MA 02138, USA}

\author{Jonathan Irwin}
\affiliation{Center for Astrophysics, Harvard \& Smithsonian, 60 Garden Street, Cambridge, MA 02138, USA}

\received{23 September, 2021}
\revised{1 March, 2022}
\accepted{9 March, 2022}

\begin{abstract}
In the canonical theory of stellar magnetic dynamo, the tachocline in partially
	convective stars serves to arrange small-scale fields, generated by
	stochastic movement of plasma into a coherent large-scale field.
	Mid-to-late M-dwarfs, which are fully convective, show more magnetic
	activity than classical magnetic dymano theory predicts. However,
	mid-to-late M-dwarfs show tight correlations between rotation and magnetic
	activity, consistent with elements of classical dynamo theory. We use data
	from Magellan Inamori Kyocera Echelle (MIKE) Spectrograph to detail the
	relation between Ca II H\&K flux and rotation period for these low-mass
	stars. We measure $R'_{HK}$ values for 53 spectroscopically identified
	M-dwarfs selected from the MEarth survey; these stars span spectral classes
	from M5.0 to M3.5 and have rotation periods ranging from hours to months.
	We present the rotation--activity relationship as traced through these
	data. We find power law and saturated regimes consistent to within one
	sigma of previously published results and observe a mass dependence in
	R'$_{HK}$.
\vspace{1cm}
\end{abstract}

\keywords{M dwarf stars (982), Stellar Activity (1580)}


\vspace{1in}
\section{Introduction} \label{sec:intro}
M-dwarfs are the most numerous stars in our galaxy. Some planet search
campaigns have focused on M-dwarfs due to the relative ease of detecting small
planets in their habitable zones \citep[e.g.][]{Nut08}; however, spun-up M-dwarfs are
more magnetically active when compared to larger and hotter stars \citep{Haw91,
Del98, Sch14}. The increase in activity may accelerate the stripping of an
orbiting planet's atmosphere \citep[e.g.][]{Owe16}, and may dramatically impact
habitability \citep{Shi16}. Therefore, it is essential to understand the
magnetic activity of M-dwarfs in order to constrain the potential habitability
and history of the planets that orbit them. Additionally, rotation and activity
may impact the detectability of hosted planets \citep[e.g.][]{Rob14,
Newton2016, Van16}.

Robust theories explaining the origin of solar-like magnetic fields exist and
have proven extensible to other regions of the main sequence \citep{Cha14}. The
classical $\alpha\Omega$ dynamo relies on differential rotation between layers
of a star to stretch a seed poloidal field into a toroidal field \citep{Par55,
Cam17}. Magnetic buoyancy causes the toroidal field to rise through the star.
During this rise, turbulent helical stretching converts the toroidal field back
into a poloidal field \citep{Par55}. Seed fields may originate from the
stochastic movement of charged particles within a star's atmospheres.

In non-fully convective stars the initial conversion of the toroidal field to a
poloidal field is believed to take place at the interface layer between the
radiative and convective regions of a star --- the tachocline \citep{Noy84,
Tom96, Dik99}. The tachocline has two key properties that allow it to play an
important role in solar type magnetic dynamos: 1), there are high shear
stresses, which have been confirmed by astroseismology \citep{Tho96}, and 2),
the density stratification between the radiative and convective zones serves to
``hold'' the newly generated toroidal fields at the tachocline for an extended
time. Over this time, the fields build in strength significantly more than they
would otherwise \citep{Par75}. This theory does not trivially extend to
mid-late M-dwarfs, as they are believed to be fully convective and consequently
do not contain a tachocline \citep{Cha97}. Moreover, fully convective M-dwarfs
are not generally expected to exhibit internal differential rotation
\citep[e.g.][]{barnes2004differential, barnes2005dependence}, though, some
models do produce it \citep{Yad13}.

Currently, there is no single accepted process that serves to build and
maintain fully convective M-dwarf magnetic fields in the same way that the
$\alpha$ and $\Omega$ processes are presently accepted in solar magnetic dynamo
theory. Three-dimensional magneto anelastic hydrodynamical simulations have
demonstrated that local fields generated by convective currents can self
organize into large scale dipolar fields. 
These models indicate that for a fully convective star to sustain a magnetic
field it must have a high degree of density stratification --- density
contrasts greater than 20 at the tachocline --- and a sufficiently large magnetic Reynolds
number\footnote{The Reynolds Number is the ratio of magnetic induction to
magnetic diffusion; consequently, a plasma with a larger magnetic Reynolds
number will  sustain a magnetic field for a longer time than a plasma with a
smaller magnetic Reynolds number.}.

An empirical relation between the rotation rate and the level of magnetic
activity has been demonstrated in late-type stars \citep{Skumanich1972, Pal81}. This is
believed to be a result of faster rotating stars exhibiting excess non-thermal
emission from the upper chromosphere or corona when compared to their slower
rotating counterparts. This excess emission is due to magnetic heating of the
upper atmosphere, driven by the underlying stellar dynamo.
\textbf{The faster a star rotates, up to some saturation threshold, the more such emission is expected.} However,
the dynamo process is not dependent solely on rotation; rather, it depends on
whether the contribution from the rotational period ($P_{rot}$) or convective
motion --- parameterized by the convective overturn time scale ($\tau_{c}$) ---
dominates the motion of a charge packet within a star. Therefore, the Rossby
Number ($Ro = P_{rot}/\tau_{c}$) is often used in place of the rotational
period as it accounts for both.

The rotation-activity relation was first discovered using the ratio of X-ray
luminosity to bolometric luminosity ($L_{X}/L_{bol}$) \citep{Pal81} and was
later demonstrated to be a more general phenomenon, observable through other
activity tracers, such as Ca II H\&K emission \citep{Vilhu1984}. This relation has
a number of important structural elements. \citet{Noy84} showed that magnetic
activity as a function of Rossby Number is well modeled as a piecewise power
law relation including a saturated and non-saturated regime. In the saturated
regime, magnetic activity is invariant to changes in Rossby Number; in the
non-saturated regime, activity decreases as Rossby Number increases. The
transition between the saturated and non-saturated regions occurs at $Ro \sim
0.1$ \citep[e.g.][]{Wri11}. Recent evidence may suggest that, instead of an
unsaturated region where activity is fully invariant to rotational period,
activity is more weakly, but still positively, correlated with rotation rate
\citep{Mamajek2008, Reiners2014, Leh20, Magaudda2020}. 

Previous studies of the Ca II H\&K rotation-activity relation
\citep[e.g.][]{Vau81, Sua15, Def17, Hou17} have focused on on spectral ranges
which both extend much earlier than M-dwarfs and which do not fully probe late
M-dwarfs. Other studies have relied on $v\sin(i)$ measurements
\citep[e.g.][]{Browning2010, Hou17}, which are not sensitive to the long
rotation periods reached by slowly rotating, inactive mid-to-late type M dwarfs
\citep[70-150 days:][]{Newton2016}. Therefore, these studies can present only
coarse constraints on the rotation activity relation in the fully convective
regime. The sample we present in this paper is focused on mid-to-late type M
dwarfs, with photometrically measured rotational periods, while maintaining of
order the same number of targets as previous studies.  Consequently, we provide
much finer constraints on the rotation-activity relation in this regime. 



We present a high resolution spectroscopic study of 53 mid-late M-dwarfs. We
measure Ca II H\&K strengths, quantified through the $R'_{HK}$ metric, which is
a bolometric flux normalized version of the Mount Wilson S-index. These
activity tracers are then used in concert with photometrically determined
rotational periods, compiled by \citet{Newton2017}, to generate a
rotation--activity relation for our sample. This paper is organized as follows:
Section \ref{sec:Observations} provides an overview of the observations and
data reduction, Section \ref{sec:Analysis} details the analysis of our data,
and Section \ref{sec:results} presents our results and how they fit within the
literature.

\section{Observations \& Data Reduction}\label{sec:Observations}
We initially selected a sample of 55 mid-late M-dwarfs from targets of
the MEarth survey \citep{Ber12} to observe. Targets were selected based on high
proper motions and availability of a previously measured photometric rotation
period, or an expectation of a measurement based on data available from
MEarth-South at the time. These rotational periods were derived photometrically
\citep[e.g.][]{Newton2016,Man16,Med20}. For star 2MASS J06022261-2019447, which
was categorized as an ``uncertain detection'' from MEarth photometry by
\citet{Newton2018}, including new data from MEarth DR10 we find a period of 95
days. This value was determined following similar methodology to \citet{Irw11}
and \citet{Newton2016,Newton2018}, and is close to the reported candidate
period of 116 days.  References for all periods are provided in the machine
readable version of Table \ref{tab:finalData}.   

High resolution spectra were collected from March to October 2017 using the
Magellan Inamori Kyocera Echelle (MIKE) spectrograh on the 6.5 meter Magellan 2
telescope at the Las Campanas Observatory in Chile. MIKE is a high resolution
double echelle spectrograph with blue and red arms. Respectively, these cover
wavelengths from 3350 - 5000 \AA\ and 4900-9500 \AA\ \citep{Ber03}. We
collected data using a 0.75x5.00" slit resulting in a resolving power of 32700.
Each science target was observed an average of four times with mean integration
times per observation ranging from 53.3 to 1500 seconds. Ca II H\&K
lines were observed over a wide range of signa-to-noise ratios, from $\sim 5$ up
to $\sim 240$ with mean and median values of 68 and 61 respectively.

We use the \texttt{CarPy} pipeline \citep{Kel00, Kel03} to reduce our blue arm
spectra. \texttt{CarPy}'s data products are wavelength calibrated, blaze
corrected, and background subtracted spectra comprising 36 orders. We shift all
resultant target spectra into the rest frame by cross correlating against a
velocity template spectrum. For the velocity template we use an observation of
Proxima Centari in our sample. This spectrum's velocity is both barycentrically
corrected, using astropy's \texttt{SkyCoord} module \citep{Ast18}, and
corrected for Proxima Centari's measured radial velocity, -22.4 km s$^{-1}$
\citep{Tor06}. Each echelle order of every other target observation is cross
correlated against the corresponding order in the template spectra using
\texttt{specutils} \texttt{template\_correlate} function \citep{Nic21}.
Velocity offsets for each order are inferred from a Gaussian fit to the
correlation vs. velocity lag function. For each target, we apply a three sigma
clip to list of echelle order velocities, visually verifying this clip removed
low S/N orders. We take the mean of the sigma-clipped velocities Finally, each
wavelength bin is shifted according to its measured velocity.

Ultimately, two targets (2MASS J16570570-0420559 and 2MASS J04102815-5336078)
had S/N ratios around the Ca II H\&K lines which were too low to be of use,
reducing the number of R'$_{HK}$ measurement we can make from 55 to 53.

\section{Analysis}\label{sec:Analysis}
\begin{figure*}[ht!]
    \centering
    \includegraphics[width=0.9\textwidth]{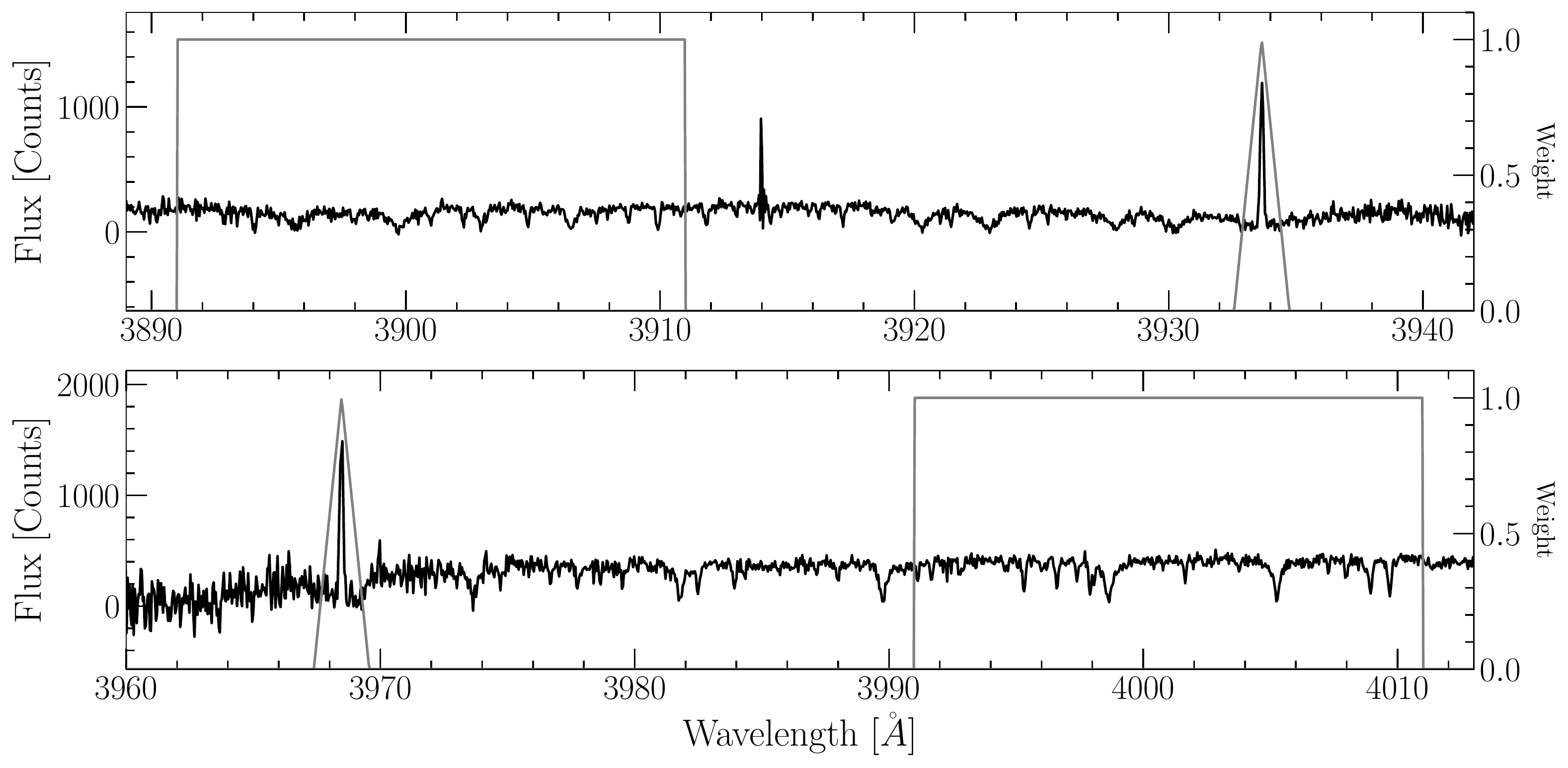}
	\caption{Spectrum of 2MASS J06105288-4324178 overplotted with the S index
	bandpasses. (top) V band and Ca II K emission line. (bottom) Ca II H
	emission line and R band. Note that the rectangular and triangular
	regions denote both the wavelength range of the band and the relative
	weight assigned to each wavelength in the band while integrating. }
    \label{fig:SindexBandpass}
\end{figure*}

Since the early 1960s, the Calcium Fraunhaufer lines have been used as
chromospheric activity tracers \citep{Wil63}. Ca II H\&K lines are observed as
a combination of a broad absorption feature originating in the upper
photosphere along with a narrow emission feature from non-thermal heating of
the upper chromosphere \citep{Catalano1983}. Specifically, the ratio between
emission in the Ca II H\&K lines and flux contributed from the photosphere is
used to define an activity metric known as the S-index \citep{Wil68}.   The
S-index increases with increasing magnetic activity. The S-index is defined as 

\begin{equation}\label{eqn:SIndex}
    S = \alpha \frac{f_{H} + f_{K}}{f_{V} + f_{R}}    
\end{equation}

\noindent where $f_{H}$  and $f_{K}$ are the integrated flux over triangular
passbands with a full width at half maximum of $1.09\text{ \AA}$ centered at
$3968.47\text{ \AA}$ and $3933.66\text{ \AA}$, respectively. The values of
$f_{V}$ and $f_{R}$ are integrated, top hat, broadband regions. They
approximate the continuum (Figure \ref{fig:SindexBandpass}) and are centered at
3901 \AA \ and 4001 \AA \ respectively, with widths of 20 \AA \ each. Finally,
$\alpha$ is a scaling factor with $\alpha = 2.4$.

Following the procedure outlined in \citet{Lov11} we use the mean flux per
wavelength interval, $\tilde{f_{i}}$, as opposed to the integrated flux over
each passband when computing the S-index. This means that for each passband,
$i$, with a blue most wavelength $\lambda_{b,i}$ and a red most wavelength
$\lambda_{r,i}$, $\tilde{f}_{i}$ is the summation of the product of flux ($f$)
and weight ($w_{i})$ over the passband.

\begin{equation}\label{eqn:meanFlux}
    \tilde{f}_{i} = \frac{\sum_{l = \lambda_{b,i}}^{\lambda_{r,i}}f(l)w_{i}(l)}{\lambda_{r,i}-\lambda_{b,i}}    
\end{equation}
\noindent where $w_{i}$ represents the triangular passband for $f_{H}$ \& $f_{K}$ and the tophat for $f_{V}$ \& $f_{R}$.

Additionally, the spectrograph used at Mount Wilson during the development of
the S-index exposed the H \& K lines for eight times longer than the continuum
of the spectra. Therefore, for a modern instrument that exposes the entire
sensor simultaneously, there will be 8 times less flux in the Ca II H\&K
passbands than the continuum passbands than for historical observations. This
additional flux is accounted for by defining a new constant $\alpha_{H}$,
defined as:

\begin{equation}
    \alpha_{H} = 8\alpha\left(\frac{1.09\text{ \AA}}{20\text{ \AA}}\right)
\end{equation}
Therefore, S-indices are calculated here not based on the historical definition
given in Equation \ref{eqn:SIndex}; rather, the slightly modified version:

\begin{equation}\label{eqn:finalSIndex}
    S = \alpha_{H}\frac{\tilde{f}_{H} + \tilde{f}_{K}}{\tilde{f}_{V} + \tilde{f}_{R}}
\end{equation}

The S-index may be used to make meaningful comparisons between stars of similar
spectral class; however, it does not account for variations in photospheric
flux and is therefore inadequate for making comparisons between stars of
different spectral classes. The $R'_{HK}$ index \citep{Middelkoop1982} is a
transformation of the S-index intended to remove the contribution of the
photosphere. 

$R'_{HK}$ introduces a bolometric correction factor, $C_{cf}$, developed by
\citet{Middelkoop1982} and later improved upon by \citet{Rutten1984}.
Calibrations of $C_{cf}$ have focused on FGK-type stars using broad band color
indices, predominately B-V. However, these FGK-type solutions do not extend to
later type stars easily as many mid-late M-dwarfs lack B-V photometry.
Consequently, $C_{cf}$ based on B-V colors were never calibrated for M-dwarfs
as many M-dwarfs lack B and V photometry. \citet{SuarezMascareno2016} provided
the first $C_{cf}$ calibrations for M-dwarfs using the more appropriate color
index of $V-K$. The calibration was later extended by \citet{Def17}, which we
adopt here. 

Generally $R'_{HK}$ is defined as

\begin{equation}\label{eqn:RpHKDef}
    R'_{HK} = K\sigma^{-1}10^{-14}C_{cf}(S-S_{phot})
\end{equation}
where K is a factor to scale surface fluxes of arbitrary units into physical
units; the current best value for K is taken from \citet{Hal07},
$K=1.07\times10^{6}\text{erg cm$^{-2}$ s$^{-1}$}$. $S_{phot}$ is the
photospheric contribution to the S-index; in the spectra this manifests as the
broad absorption feature wherein the narrow Ca II H\&K emission resides.
$\sigma$ is the Stephan-Boltzmann constant. If we define 

\begin{equation}
    R_{phot}\equiv K\sigma^{-1}10^{-14}C_{cf}S_{phot}
\end{equation}
then we may write $R'_{HK}$ as 

\begin{equation}\label{eqn:RpHKFinal}
    R'_{HK} = K\sigma^{-1}10^{-14}C_{cf}S - R_{phot}.
\end{equation}

We use the color calibrated coefficients for $\log_{10}(C_{cf})$ and
$\log_{10}(R_{phot})$ presented in Table 1 of \citet{Def17}.

We estimate the uncertainty of $R'_{HK}$ as the standard deviation of a
distribution of $R'_{HK}$ measurements from 5000 Monte Carlo tests. For each
science target we offset the flux value at each wavelength bin by an amount
sampled from a normal distribution. The standard deviation of this normal
distribution is equal to the estimated error at each wavelength bin. These
errors are calculated at reduction time by the pipeline. The R$'_{HK}$
uncertainty varies drastically with signal-to-noise; targets with
signal-to-noise ratios $\sim 5$ have typical uncertainties of a few percent
whereas targets with signal-to-noise ratios $\sim 100$ have typical
uncertainties of a few tenth of a percent.

\subsection{Rotation and Rossby Number}
The goal of this work is to constrain the rotation activity relation;
therefore, in addition to the measured $R'_{HK}$ value, we also need the
rotation of the star. As mentioned, one of the selection criteria for targets
was that their rotation periods were already measured; however, ultimately
6 of the 53 targets with acceptable S/N did not have well constrained
rotational periods. We therefore only use the remaining 47 targets to fit the
rotation-activity relation. 


In order to make the most meaningful comparison possible we transform rotation
period into Rossby Number . This transformation was done using the convective
overturn timescale, $\tau_{c}$, such that the Rossby Number, $Ro =
P_{rot}/\tau_{c}$ . To first order $\tau_{c}$ can be approximated as $70$ days
for fully-convective M-dwarfs \citep{Pizzolato2000}. However, \citet{Wri18}
Equation (5) presents an empirically calibrated expression for $\tau_{c}$. This
calibration is derived by fitting the convective overturn timescale as a
function of color index, in order to minimize the horizontal offset between
stars of different mass in the rotation-activity relationship.  The calibration
from \citet{Wri18} that we use to find convective overturn timescales and
subsequently Rossby numbers is:

\begin{equation}\label{eqn:convectiveOverturn}
    \log_{10}(\tau_{c}) = (0.64\pm0.12)+(0.25\pm0.08)(V-K)
\end{equation}
We adopt symmetric errors for the parameters of Equation
\ref{eqn:convectiveOverturn} equal to the larger of the two anti-symmetric
errors presented in \citet{Wri18} Equation 5.

\section{Rotation--Activity Relation}\label{sec:results}
\begin{figure*}
    \centering
    \includegraphics[width=0.9\textwidth]{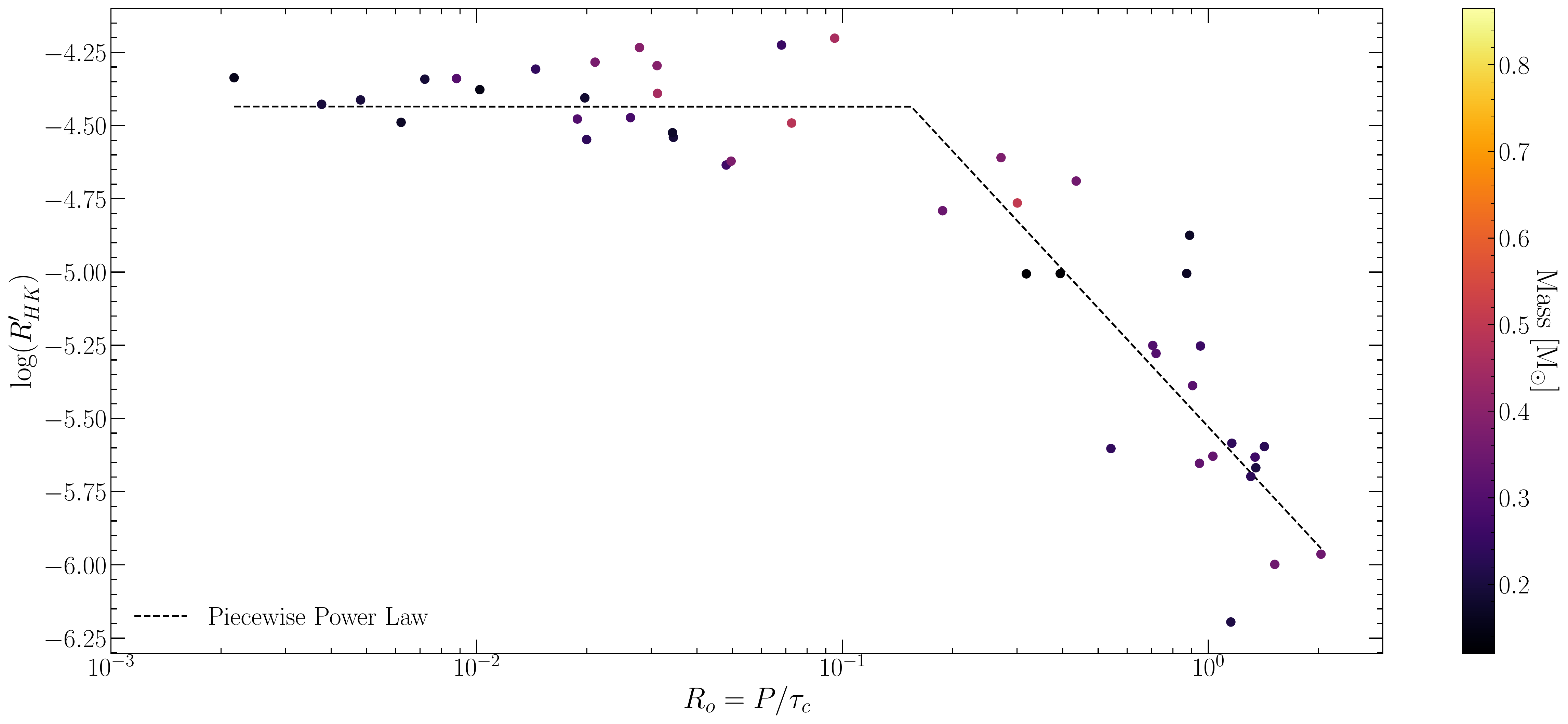}
	\caption{Rotation activity relation from this work. The color axis gives
	each stars mass. The dashed line is the best fit to our data set.}
    \label{fig:RpHKvsRossbySelf}
\end{figure*}

\begin{figure*}
    \centering
    \includegraphics[width=0.9\textwidth]{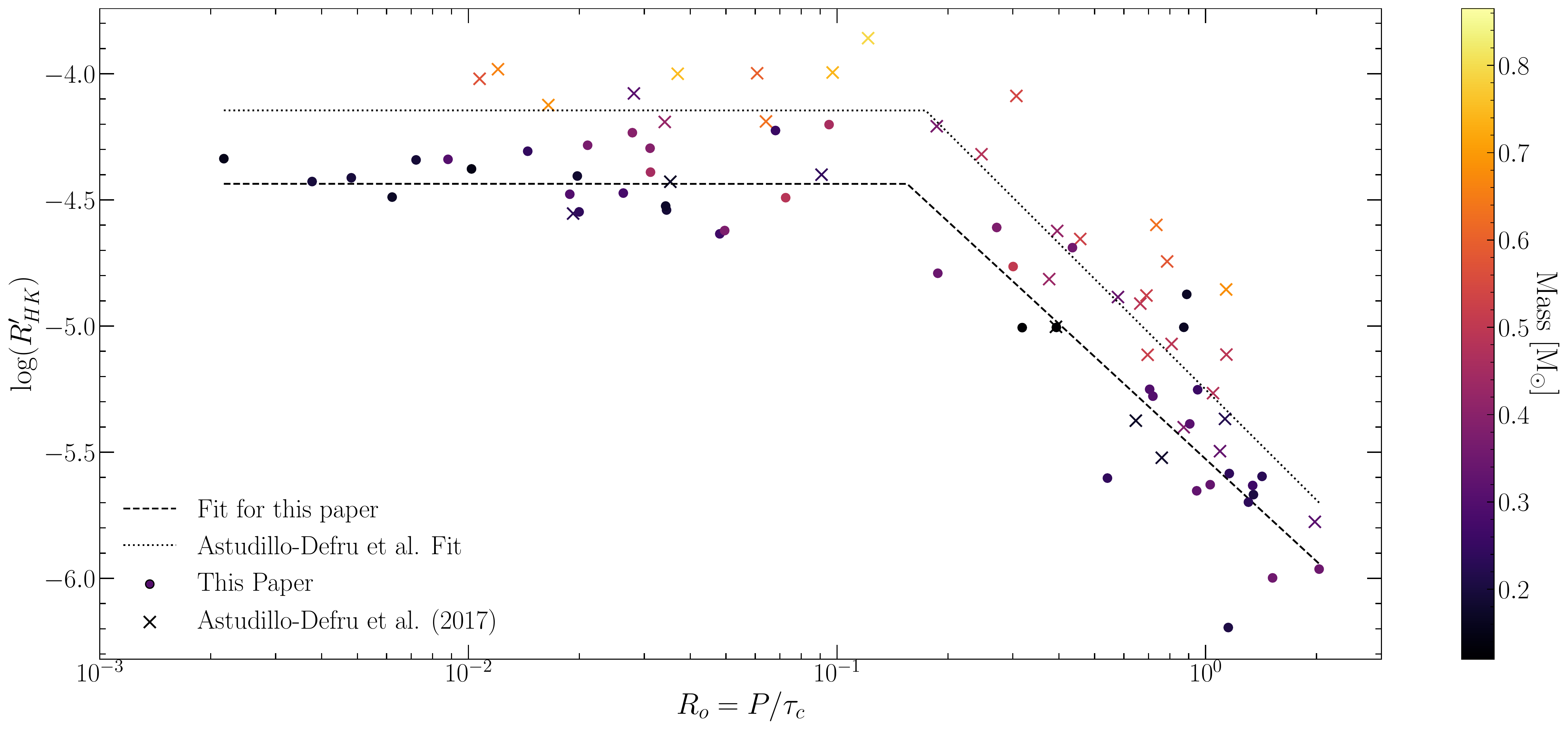}
	\caption{Rotation activity relation for both our work and \citet{Def17}.
	The dotted line is the best fit to the re-derived rotation-activity
	relation from \citet{Def17}.  Note that targets from \citet{Def17} are
	systematically higher than targets presented here as a consequence of the
	range in mass probed by the samples.}
    \label{fig:RpHKvsRossbyDef}
\end{figure*}

We show our rotation-activity relation in Figures
\ref{fig:RpHKvsRossbySelf} \& \ref{fig:RpHKvsRossbyDef}. Note that
errors are shown in both figures; however, they render smaller than the data
point size. Ca II H\&K is also known to be time variable
\citep[e.g.][]{Baroch2020,Perdelwitz2021}, which is not captured in our
single-epoch data. There is one target cut off by the domain of this graph,
2MASS J10252645+0512391. This target has a measured vsini of $59.5\pm2.1$ km
s$^{-1}$ \citep{Kesseli2018} and is therefore quite rotationally broadened, which
is known to affect $R'_{HK}$ measurements \citep[figure 8]{Schroder2009}. The
data used to generate this figure is given in Table \ref{tab:finalData}. Table
\ref{tab:finalData} includes uncertainties, the R'$_{HK}$ measurements for
stars which did not have photometrically derived rotational periods in MEarth,
and data for 2MASS J10252645+0512391

We find a rotation activity relationship qualitatively similar to that
presented in \citet{Def17}. Our rotation activity relationship exhibits both
the expected saturated and unsaturated regimes --- the flat region at $Ro <
Ro_{s}$ and the sloped region at $Ro \geq Ro_{s}$ respectively. We fit the
rotation activity relation given in Equation \ref{eqn:fitEqn} to our data using
Markov Chain Monte Carlo (MCMC), implemented in \texttt{pymc}
\citep{Salvatier2016}. 

{\scriptsize
    \begin{equation}\label{eqn:fitEqn}
        \log(R'_{HK}) = \begin{cases}
            \log(R_{s}) & Ro < Ro_{s} \\
            k\log(Ro) + \log(R_{s}) - k\log(Ro_{s}) & Ro \geq Ro_{s}
        \end{cases}
    \end{equation}
}

\noindent $Ro_{s}$ is the Rossby number cutoff between the saturated and
unsaturated regime. $R_{s}$ is the maximum, saturated, value of $R'_{HK}$ and
$k$ is the index of the power law when $Ro \geq Ro_{s}$. Due to the
issues measuring $R'_{HK}$ for high vsini targets discussed above, we exclude
2MASS J10252645+0512391 from this fit. All logarithms are base ten unless
another base is explicitly given.

\begin{table*}[]
    \centering
    \setlength{\tabcolsep}{4pt}
	\renewcommand{\arraystretch}{0.9} 
    \begin{tabular}{lccccccccc}
\hline
2MASS ID & Mass & $Ro$ & $\log(R'_{HK})$ & $\log(R'_{HK})_{err}$ & $V_{mag}$ & $V-K$ & prot & $r_{prot}$ & Estimate \\
 & $\mathrm{M_{\odot}}$ &  &  &  & $\mathrm{mag}$ & $\mathrm{mag}$ & $\mathrm{d}$ &  &  \\
\hline
\hline
06000351+0242236 & 0.24 & 0.020 & -4.5475 & 0.0021 & 11.31 & 5.268 & 1.809 & 2016ApJ...821...93N & False \\
02125458+0000167 & 0.27 & 0.048 & -4.6345 & 0.0014 & 13.58 & 5.412 & 4.732 & 2016ApJ...821...93N & False \\
01124752+0154395 & 0.28 & 0.026 & -4.4729 & 0.0017 & 14.009 & 5.240 & 2.346 & 2016ApJ...821...93N & False \\
10252645+0512391 & 0.11 & 0.000 & -4.9707 & 0.0380 & 18.11 & 7.322 & 0.102 & 2016ApJ...821...93N & False \\
05015746-0656459 & 0.17 & 0.873 & -5.0049 & 0.0028 & 12.2 & 5.464 & 88.500 & 2012AcA....62...67K & False \\
06022261-2019447 & 0.23 & 1.307 & -5.6980 & 0.0192 & 13.26 & 4.886 & 95.000 & This Work & False \\
06105288-4324178 & 0.30 & 0.705 & -5.2507 & 0.0139 & 12.28 & 4.968 & 53.736 & 2018AJ....156..217N & False \\
09442373-7358382 & 0.24 & 0.542 & -5.6026 & 0.0147 & 15.17 & 5.795 & 66.447 & 2018AJ....156..217N & False \\
14211512-0107199 & 0.24 & 1.160 & -5.5846 & 0.0125 & 13.12 & 5.027 & 91.426 & 2018AJ....156..217N & False \\
14294291-6240465 & 0.12 & 0.394 & -5.0053 & 0.0014 & 11.13 & 6.746 & 83.500 & 1998AJ....116..429B & False \\
16352464-2718533 & 0.23 & 1.423 & -5.5959 & 0.0108 & 14.18 & 5.182 & 122.656 & 2018AJ....156..217N & False \\
16570570-0420559 & 0.24 & 0.014 & -4.3071 & 0.0014 & 12.25 & 5.130 & 1.212 & 2012AcA....62...67K & False \\
02004725-1021209 & 0.34 & 0.188 & -4.7907 & 0.0026 & 14.118 & 5.026 & 14.793 & 2018AJ....156..217N & False \\
18494929-2350101 & 0.18 & 0.034 & -4.5243 & 0.0015 & 10.5 & 5.130 & 2.869 & 2007AcA....57..149K & False \\
20035892-0807472 & 0.33 & 0.946 & -5.6530 & 0.0077 & 13.54 & 5.254 & 84.991 & 2018AJ....156..217N & False \\
21390081-2409280 & 0.21 & 1.152 & -6.1949 & 0.0190 & 13.45 & 5.091 & 94.254 & 2018AJ....156..217N & False \\
23071524-2307533 & 0.30 & 0.720 & -5.2780 & 0.0077 & 13.587 & 4.849 & 51.204 & 2018AJ....156..217N & False \\
00094508-4201396 & 0.30 & 0.009 & -4.3392 & 0.0018 & 13.62 & 5.397 & 0.859 & 2018AJ....156..217N & False \\
00310412-7201061 & 0.31 & 0.906 & -5.3879 & 0.0074 & 13.69 & 5.245 & 80.969 & 2018AJ....156..217N & False \\
01040695-6522272 & 0.17 & 0.006 & -4.4889 & 0.0024 & 13.98 & 5.448 & 0.624 & 2018AJ....156..217N & False \\
02014384-1017295 & 0.19 & 0.034 & -4.5400 & 0.0022 & 14.473 & 5.284 & 3.152 & 2018AJ....156..217N & False \\
03100305-2341308 & 0.40 & 0.028 & -4.2336 & 0.0017 & 13.502 & 4.935 & 2.083 & 2018AJ....156..217N & False \\
03205178-6351524 & 0.33 & 1.029 & -5.6288 & 0.0096 & 13.433 & 5.238 & 91.622 & 2018AJ....156..217N & False \\
07401183-4257406 & 0.15 & 0.002 & -4.3365 & 0.0022 & 13.81 & 6.042 & 0.307 & 2018AJ....156..217N & False \\
08184619-4806172 & 0.37 & 0.021 & -4.2834 & 0.0025 & 14.37 & 5.019 & 1.653 & 2018AJ....156..217N & False \\
08443891-4805218 & 0.20 & 1.348 & -5.6682 & 0.0067 & 13.932 & 5.370 & 129.513 & 2018AJ....156..217N & False \\
09342791-2643267 & 0.19 & 0.007 & -4.3415 & 0.0025 & 13.992 & 5.373 & 0.694 & 2018AJ....156..217N & False \\
09524176-1536137 & 0.26 & 1.342 & -5.6319 & 0.0110 & 13.43 & 4.923 & 99.662 & 2018AJ....156..217N & False \\
11075025-3421003 & 0.25 & 0.068 & -4.2250 & 0.0032 & 15.04 & 5.633 & 7.611 & 2018AJ....156..217N & False \\
11575352-2349007 & 0.39 & 0.031 & -4.2952 & 0.0026 & 14.77 & 5.415 & 3.067 & 2018AJ....156..217N & False \\
12102834-1310234 & 0.36 & 0.435 & -4.6892 & 0.0029 & 13.83 & 5.418 & 42.985 & 2018AJ....156..217N & False \\
12440075-1110302 & 0.18 & 0.020 & -4.4053 & 0.0033 & 14.22 & 5.546 & 2.099 & 2018AJ....156..217N & False \\
13442092-2618350 & 0.35 & 2.032 & -5.9634 & 0.0253 & 13.253 & 4.968 & 154.885 & 2018AJ....156..217N & False \\
14253413-1148515 & 0.51 & 0.301 & -4.7641 & 0.0030 & 13.512 & 5.121 & 25.012 & 2018AJ....156..217N & False \\
14340491-1824106 & 0.38 & 0.271 & -4.6093 & 0.0038 & 14.346 & 5.638 & 30.396 & 2018AJ....156..217N & False \\
15154371-0725208 & 0.38 & 0.050 & -4.6214 & 0.0023 & 12.93 & 5.224 & 4.379 & 2018AJ....156..217N & False \\
15290145-0612461 & 0.46 & 0.095 & -4.2015 & 0.0017 & 14.011 & 5.230 & 8.434 & 2018AJ....156..217N & False \\
16204186-2005139 & 0.45 & 0.031 & -4.3900 & 0.0035 & 13.68 & 5.261 & 2.814 & 2018AJ....156..217N & False \\
16475517-6509116 & 0.17 & 0.889 & -4.8744 & 0.0045 & 13.98 & 5.101 & 73.142 & 2018AJ....156..217N & False \\
20091824-0113377 & 0.15 & 0.010 & -4.3772 & 0.0023 & 14.47 & 5.958 & 1.374 & 2018AJ....156..217N & False \\
20273733-5452592 & 0.35 & 1.520 & -5.9982 & 0.0181 & 13.18 & 5.259 & 136.924 & 2018AJ....156..217N & False \\
20444800-1453208 & 0.49 & 0.073 & -4.4912 & 0.0023 & 14.445 & 5.305 & 6.715 & 2018AJ....156..217N & False \\
15404341-5101357 & 0.10 & 0.318 & -5.0062 & 0.0081 & 15.26 & 7.317 & 93.702 & 2018AJ....156..217N & False \\
22480446-2422075 & 0.20 & 0.005 & -4.4123 & 0.0016 & 12.59 & 5.384 & 0.466 & 2013AJ....146..154M & False \\
06393742-2101333 & 0.26 & 0.952 & -5.2524 & 0.0069 & 12.77 & 5.120 & 79.152 & 2018AJ....156..217N & False \\
04130560+1514520 & 0.30 & 0.019 & -4.4775 & 0.0088 & 15.881 & 5.437 & 1.881 & 2016ApJ...818..46M & False \\
02411510-0432177 & 0.20 & 0.004 & -4.4272 & 0.0016 & 13.79 & 5.544 & 0.400 & 2020ApJ...905..107M & False \\
11381671-7721484 & 0.12 & 0.958 & -5.5015 & 0.0369 & 14.78 & 6.259 & 153.506 & This Work & True \\
12384914-3822527 & 0.15 & 2.527 & -6.0690 & 0.0156 & 12.75 & 5.364 & 241.913 & This Work & True \\
13464102-5830117 & 0.48 & 1.340 & -5.6977 & 0.0146 &  &  & 65.017 & This Work & True \\
15165576-0037116 & 0.31 & 0.157 & -4.0704 & 0.0024 & 14.469 & 5.364 & 15.028 & This Work & True \\
19204795-4533283 & 0.18 & 1.706 & -5.8392 & 0.0091 & 12.25 & 5.405 & 167.225 & This Work & True \\
21362532-4401005 & 0.20 & 1.886 & -5.8978 & 0.0168 & 14.14 & 5.610 & 207.983 & This Work & True \\
\hline
\end{tabular}

	\caption{Calculated Rossby Numbers and $R'_{HK}$ values. All circular data
	points in Figures \ref{fig:RpHKvsRossbySelf} \& \ref{fig:RpHKvsRossbyDef}
	are present in this table. Masses are taken from the MEarth database. A
	machine readable version of this table is available}
    \label{tab:finalData}
\end{table*}
\begin{figure*}
    \centering
    \includegraphics[width=0.9\textwidth]{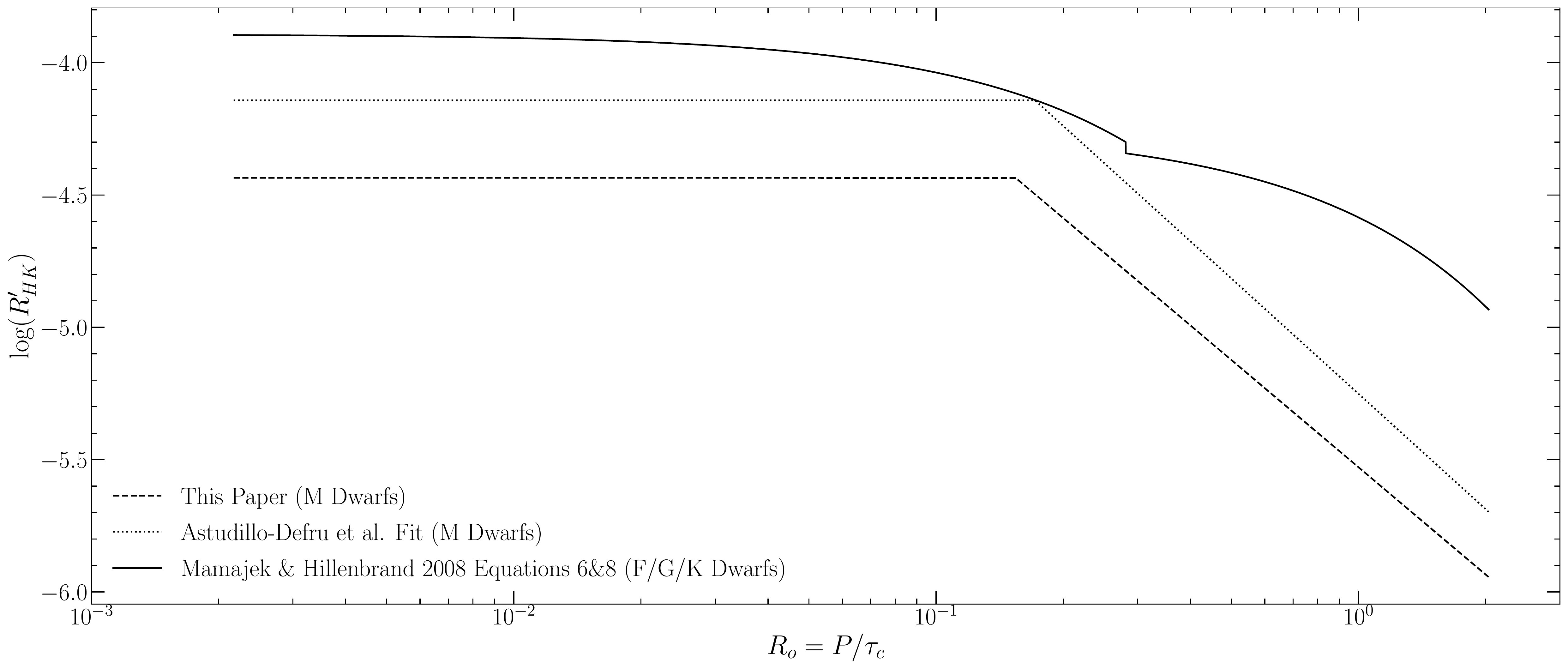}
	\caption{Derived rotation-activity curves from this work, \citet{Def17} and
	\citet{Mamajek2008}. Note both that \citet{Mamajek2008} focuses their work
	on earlier spectral classes and fits the rotation activity relation in
	linear space.}
    \label{fig:RpHKvsRossbyFits}
\end{figure*}

We find best fit parameters with one $\sigma$ errors:
\begin{itemize}
    \item $k = -1.347\pm 0.203$
    \item $Ro_{s} =  0.155\pm0.045$
    \item $\log(R_{s}) = -4.436\pm0.048$

\end{itemize}
A comparison of the rotation activity derived in this work to those
from both \citet{Def17} and \citet{Mamajek2008} is presented in Figure
\ref{fig:RpHKvsRossbyFits}. For the 6 targets which do not have measured
rotational periods we include an estimate of $Ro$ and $p_{rot}$ in the machine
readable version of Table \ref{tab:finalData}. The convective overturn
timescale for one of these 6 targets (2MASS J13464102-5830117) can not be
inferred via Equation \ref{eqn:convectiveOverturn} as it lacks a V-K color
measurement. Instead, we infer $\tau_{c}$ via \citet{Wri18} Equation 6 (this
paper Equation \ref{eqn:ConvectiveOverturnTimeMass}) using mass. Similar to our
manner of inferring $\tau_{c}$ via color, when inferring $\tau_c$ via mass, we
adopt the larger of the two antisymmetric errors from \citet{Wri18}.

{\scriptsize
	\begin{equation}\label{eqn:ConvectiveOverturnTimeMass}
		\log_{10}(\tau_{c}) = 2.33\pm0.06 - 1.5\pm0.21\left(M/M_{\odot}\right) + 0.31\pm0.17\left(M/M_{\odot}\right)^{2}
	\end{equation}
}

Note that $R'_{HK}$ for one of six of these targets (2MASS
J15165576-0037116) is consistent to within 1$\sigma$ of the saturated value;
therefore, the reported $Ro$ for this target should only be taken as an upper
bound. The remaining five targets have measured $R'_{HK}$ values consistent
with the unsaturated regime. Estimated periods are consistent with previous
constraints. Of the six stars, two were listed as non-detections in
\citet{Newton2018}, and the remaining four as uncertain (possible) detections.
Of the four classed as uncertain, 2MASS 12384914-3822527 and 2MASS
19204795-4533283 have candidate periods $>100$ days and non-detections of
H-alpha emission \citep{Hawley96}. These two stars and the two non-detections
have Ca II H\&K activity levels suggesting very long periods. 2MASS
13464102-5830117 has a candidate period of 45 days, and 2MASS 15165576-0037116
of 0.8 days, both consistent with their higher levels of Ca II H\&K emission.

As a test of the proposed weak correlation between activity and rotation in the
``saturated'' regime seen in some works \citep{Mamajek2008,
Reiners2014, Leh20, Med20} --- though not in others \citep{Wri11, Nunez2015,
Newton2017} ---   we fit a second model whose power law index is allowed to
vary at $Ro < Ro_{s}$. We find a saturated regime power law index of
$-0.052\pm0.117$, consistent with 0 to within 1$\sigma$. Moreover,
all other parameter for this model are consistent to within one $\sigma$ of the
nominal  parameters for the model where the index is constrained to 0 below
$Ro=Ro_{s}$. We can constrain the slope in the saturated
regime to be between -0.363 and 0.259 at the $3\sigma$ confidence level.
Ultimately, we adopt the most standard activity interpretation, a
fully-saturated regime at $Ro < Ro_{s}$. 

We investigate whether our lack of detection of a slope for $Ro <
Ro_{s}$ is due to the limited number of observations in that region when
compared to other works \citep[e.g.][93 targets $Ro < Ro_{s}$]{Med20} through
injection and recovery tests. We inject, fake, rotation-activity measurements
into the saturated regime with an a priori slope of -0.13 --- the same as in
\citeauthor{Med20}. These fake data are given a standard deviation equal to the
standard deviation of our residuals ($12\%$). We preform the same MCMC model
fitting to this new data set as was done with the original dataset multiple
times, each with progressively more injected data, until we can detect the
injected slope to the three sigma confidence level. Ultimately, we need more
than 65 data points --- 43 more than we observed in the saturated regime --- to
consistently recover this slope. Therefore, given the spread of our data we
cannot detect slopes on the order of what has previously been reported in the
literature.

We observe a gap in rotational period over a comparable range to the
one presented in \citet{Newton2016} Figure 2. Namely, that M-dwarfs are
preferentially observed as either fast or slow rotators, with a seeming lack of
stars existing at mid rotational periods. This period gap manifests in the
Rossby Number and can be seen in Figure \ref{fig:RpHKvsRossbyDef} as a lack of
our targets near to the knee-point in the fit. This period gap likely
corresponds to that seen by \citet{Browning2010}, who found a paucity of M
dwarfs at intermediate activity levels in Ca II H\&K and note the similarity to
the Vaughn-Preston gap established in higher mass stars \citep{vaughan1980}.
\citet{Magaudda2020} also identify a double-gap in x-ray activity for stars in
the unsaturated regime; it is not clear that the gap we see is related. As a
consequence of this period gap, there exists a degeneracy in our data
between moving the knee-point and allowing the activity level to vary in the
saturated regime.  In the following, we adopt the model of a fully saturated
regime.

We wish to compare our best fit parameters to those derived in \citet{Def17};
however, the authors of that paper do not fit the knee-point of the
rotation-activity relation. They select the canonical value for the rotational
period separating the saturated regime from the unsaturated regime ($P_{rot,s}
= 10$ days) and use a fixed convective overturn timescale ($\tau_{c} = 70$
days). To make our comparison more meaningful we use the $P_{rot}$ and $V-K$
colors presented in \citet{Def17} to re-derive $Ro$ values using $\tau_{c}$
\citep{Wri18}. Doing this for all targets presented in \citet{Def17} Table 3
and fitting the same piecewise power law as before, we find best fit parameters
of $Ro_{s} = 0.17\pm0.04$, $\log(R_{s}) = -4.140\pm0.067$, and
$k=-1.43\pm0.21$. Compared to the best fit parameters for our data, $Ro_{s}$
and the unsaturated regime's index, $k$, are consistent to within one sigma,
while the saturated value, $R_{s}$, differs. 

\begin{figure}
    \centering
    \includegraphics[width=0.47\textwidth]{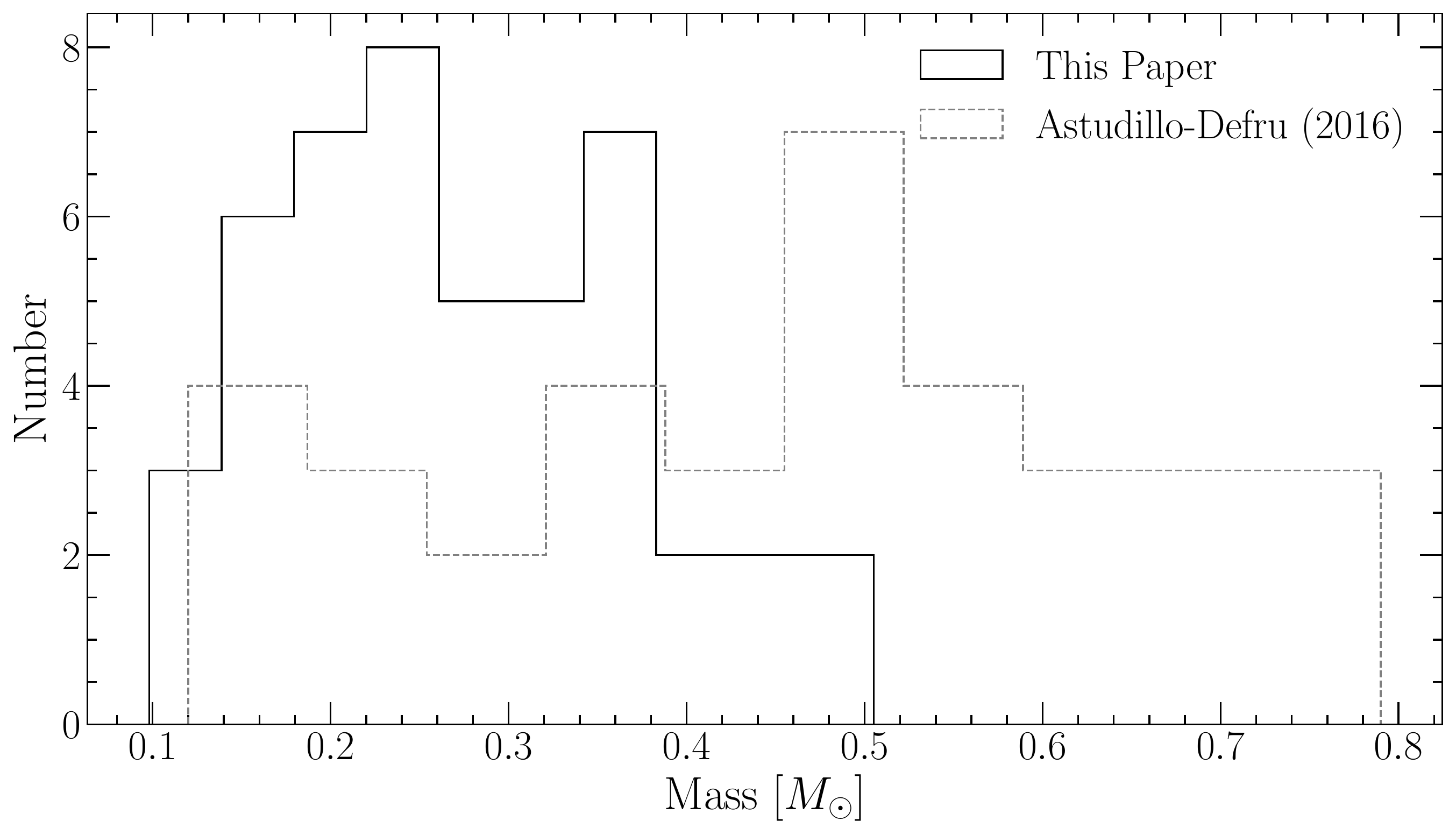}
	\caption{Distribution of masses between our sample and the sample presented
	in \citet{Def17}. Note how the two studies have approximately the same
	sample sizes; however, our sample is more tightly concentrated at lower
	masses \textbackslash later spectral classes.}
    \label{fig:massDistribution}
\end{figure}

The mass ranges of our respective samples explain the differences in saturation
values between our work and that of \citet{Def17}. Our work focuses on
mid-to-late M-dwarfs and includes no stars above a mass of $0.5$ M$_{\odot}$
(Figure \ref{fig:massDistribution}).  The strength of Ca II H\&K
emission is known to decrease as stellar mass decreases \citep{Schrijver1987,
Rauscher2006, Hou17}. As \citet{Rauscher2006} note, this is the opposite as the
trend seen in H-alpha; the latter primarily reflects the increasing length of
time that lower M dwarfs remain active and rapidly rotating \citep{West2015,
Newton2016}.

A mass dependence can be seen in Figure 10 in \citet{Def17}, consistent
with expectations from the literature. If we clip the data from \citet{Def17}
Table 3 to the same mass range as our data-set ($M_{*} < 0.5M_{\odot}$) and fit
the same function as above, we find that all best fit parameters are consistent
to within one sigma between the two data-sets. 

We also compare our best fit $Ro_{s}$ to both those derived in
\citet{Newton2017} using $H_{\alpha}$ as an activity measure and those derived
in \citep{ Wri18, Magaudda2020} using $L_{X}/L_{bol}$ as an activity measure.
Works using $L_{X}/L_{bol}$ identify a similar, yet not consistent to within
one sigma result for $Ro_{s}$; while, the value of $k$ we find here is
consistent between all four works. Therefore, we find similar results not only
to other work using the same activity tracer, but also a power-law slope that
is consistent with work using different tracers.

\section{Conclusions}\label{sec:conclusions}
In this work we have approximately doubled the number of M-dwarfs with both
empirically measured $R'_{HK}$ with $M_{*} < 0.5 M_{\odot}$. This has enabled
us to more precisely constrain the rotation-activity relation. This
relationship is consistent with other measurements using $R'_{HK}$, and
$L_{X}$/$L_{bol}$; our data does not require a slope in the saturated regime.
Finally, we identify a mass dependence in the activity level of the saturated
regime, consistent with trends seen in more massive stars in previous works.

\acknowledgments{
This work has made use of the NASA Astrophysical Data System. We thank Dan
Kelson for assistance with the \texttt{CarPy} data reduction suite. Additionally, we
would like to thank Zach Berta-Thompson, Brian Chaboyer, and John Thorstensen
for their support. We acknowledge support for an NSF AAPF awarded to Elisabeth
R. Newton (award \# 1602597). This material is based upon work supported by the
National Science Foundation under grant AST-1616624, and by a grant from the
John Templeton Foundation. The opinions expressed in this publication are those
of the authors and do not necessarily reflect the views of the John Templeton
Foundation. N.M. was supported by the National Science Foundation Graduate
Research Fellowship Program, under NSF grant No. DGE1745303. N.M. thanks the
LSSTC Data Science Fellowship Program; his time as a Fellow has greatly
benefited this work.}

\acknowledgments

\bibliography{ms}

\end{document}